# Ultraviolet luminescence of polycyclic aromatic hydrocarbons in partially consolidated sol-gel silica glasses.

Linards Skuja, Madara Leimane, Ivita Bite, Donats Millers, Aleksejs Zolotarjovs, Virginija Vitola, Krisjanis Smits

Institute of Solid State Physics, University of Latvia

8 Kengaraga str., Riga, LV1063, Latvia

## Abstract

Photoluminescence (PL) and Raman spectra of sol-gel-derived silica glasses were studied in course of the consolidation of xerogel to dense glass. A number of ultraviolet PL bands in 3 eV – 4 eV region were found, which have not been observed in pure silica glasses not exposed to carbon compounds. Such PL bands, some of which exhibit distinct vibronic structures at room temperature, have been persistently reported in various high-surface porous or nanoparticle forms of $SiO_2$. Based on their spectral shapes and decay kinetics properties, the structured PL bands in the 3.7-3.8 eV and 3.1-3.3 eV regions are assigned to polycyclic hydrocarbons, naphthalene and pyrene, respectively, which are formed from ethanol molecules created by hydrolysis of tetraethoxysilane. These data support the hypothesis that PL activity of carbon-doped silica may be related to embedded aromatic carbon groups.

# 1 Introduction

Sol-gel synthesis of pure $SiO_2$ (silica) glass has received much attention in the past decades [1]. Compared to industrial gas-phase or melt-quenching ways of synthesis, it offers a large number of flexible choices in varying the glass morphology and introduction of dopants, and requires lower processing temperatures. By careful tuning of the synthesis procedures, high purity sol-gel silica glasses are obtained ( e.g., [2,3]), having vacuum ultraviolet optical transmission comparable to the best industrial glasses produced from $SiCl_4$. In the case of chloride-based synthesis, trace amounts of Cl can remain in glass, worsening its optical properties. Chlorine is then present as network bound Si-Cl groups, or as interstitial molecules of $Cl_2$ and Cl oxides [4].

By analogy, in the case of silica glasses synthesized from silicon-organic raw materials, the presence of traces of carbon in some form can be expected. Materials with high concentrations of carbon, silicon oxycarbide glasses [5] are intensively investigated due to their applications as low-κ dielectrics [6] in microelectronics. To study low concentrations of defects/impurities, photoluminescence (PL) spectroscopy is advantageous due to its high sensitivity. There are relatively few studies, e.g., ref. [7], of optical properties of low-concentration C impurities and their atomic/molecular structure in bulk silica glasses. In contrast, PL properties of organic-modified $SiO_2$ nanoparticles and high-surface porous $SiO_2$ are intensively studied, mostly in the contexts of biomedical applications and carbon nanodots (see, e.g., references in paper [8]). Numerous new PL bands, which are not observed in chloride-derived silica glass, have been reported in the near infrared to UV spectral ranges. Unfortunately, PL spectra by themselves often provide little information on the atomic structure of the emitting centers, therefore only in few cases, e.g. for phenyl-groups grafted on silica surface [8], a more definite assignment of carbon doping-related PL bands could have been made.

The aim of the present study was to investigate the new PL bands in sol-gel silica, emerging during the consolidation of the porous xerogel to dense glass and to establish their origin. Several of the observed most intense PL bands have highly-characteristic spectral shapes, which allow to assign them with a reasonable confidence to embedded naphthalene and pyrene molecules.

## 2 Experiment

### 2.1 Samples

The $SiO_2$ glass samples were prepared from tetraethoxysilane (TEOS) by the "phase – separating sol" method adapted from ref. [9] with gelation and drying process modified with an aim to obtain crack-free $SiO_2$ glass. To obtain phase-separated sol, ammonium acetate (AcONH4) was added after partial hydrolysis of TEOS to increase pH and slow down the TEOS hydrolysis rate. 3 separate groups of dried gel (xerogel) samples were thermally annealed in air, in flowing $O_2$ or $H_2$. Within each group, 6 samples were annealed at 6 different temperatures, 200 °C to 1200 °C in steps of 200 °C by heating with rate 5 °C/min and holding at the respective temperature for 2h. Details of the sample preparation are given in Appendix.

### 2.2 Measurements

Raman spectra were measured in backscattering configuration by Andor Shamrock 303 spectrograph with 1200 l/mm grating and cooled silicon CCD, using 532.08 nm 200mW continuous wave (CW) excitation. The spectral resolution was 0.7 nm ($\approx$20 cm$^{-1}$) FWHM (full width at half maximum). Interference by sample background luminescence was mitigated by its partial photobleaching; the Raman spectrum was obtained as the non-bleachable component of the as-measured composite PL-Raman spectrum.

CW-mode PL emission and excitation (PLE) spectra were measured by Edinburgh FLS1000 scanning spectrometer with Xe-lamp as light source and double monochromators used in both excitation and emission channels for efficient stray light suppression. Both PL and PLE spectra were corrected for the spectral variations of monochromator dispersion, grating/detector efficiency and excitation lamp light intensity. The spectral resolution of PL and PLE spectra was between 0.3 nm and 2 nm; bandwidths of PL excitation or PLE emission monitoring wavelengths were 2 to 6 nm.

Time-resolved luminescence spectra were obtained using excitation by 4th harmonic ($\lambda$=266 nm) of low-power (1mW@6400 pulses per second) Nd-YAG laser (Crylas FQSS266-Q) with pulse duration 0.9 ns (FWHM). Spectra were recorded by Andor Kymera 328i 300 mm spectrograph with 150 l/mm grating and spectral resolution 1.9 nm FWHM and iStar DH320T-18U-E3 image-intensified CCD with programmable gate width and delay. Because of a large jitter ($\approx$500 ns) between the fire pulse sent to the laser and the actual light pulse, only passive triggering scheme could be used, where the camera is triggered by the laser pulse sensed by a fast photodiode. Due to the image-intensifier inherent start-up delay (~35 ns) this scheme was usable only to register PL components longer than 35 ns. The variable delays indicated in the Results section include this trigger delay. To evaluate the contributions of shorter PL components, gate widths larger than the laser pulse repetition period were used, thus including the next pulse in the registered PL signal.

Luminescence decay kinetics was roughly estimated using intensities of the time-resolved spectra at 2 different delays. More accurate decay curve was measured using the same 266 nm excitation laser source, the emission filtering through double monochromator of FLS1000 spectrometer and multichannel photon counter Edinburgh TCC2 with channel width 10 ns.

Optical transmission spectra were estimated using Ocean Optics DH2000 fiber-optics light source, reflective beam collimators and StellarNet BLK-C CCD mini-spectrometer with resolution 1.5 nm.

All measurements were performed at room temperature.

# 3 Results

## 3.1 Optical absorption and Raman spectra

The samples of thickness ≈2 mm, which were annealed at temperatures up to 1000 °C and under different atmospheres - air, $O_2$, $H_2$, had generally transparent, glass-like appearance. $H_2$-, and air-heated samples had brownish tint and $O_2$-heated samples were visually perfectly clear and colorless. The sample surfaces were not flat and uniform, preventing accurate optical transmission measurements. The obtained spectra (Fig.1), however, allow to conclude that there are no distinct strong optical absorption bands up to 4.5 eV and that 800°- annealed samples maintain some UV transparency up to 5 eV (down to 250 nm). Transmission measurements of samples annealed at higher T's were not possible since some of them cracked in small pieces or fumed ($H_2$-treated sample at 1200 °C).

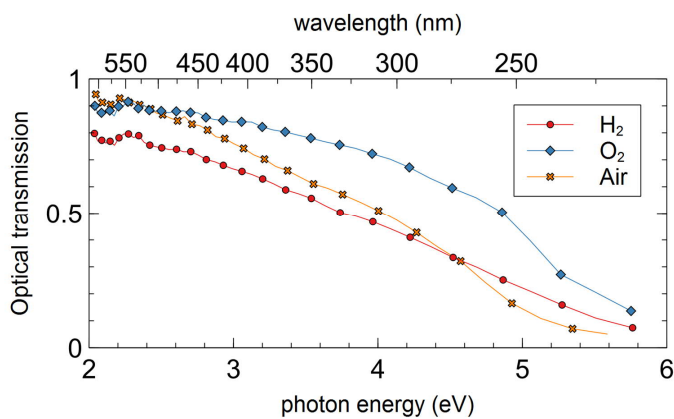

Fig.1. Optical transmission spectra of samples annealed at 800 °C under different atmospheres. Sample thickness ≈2mm.

A typical evolution of Raman spectra with increasing the annealing temperature of xerogel is depicted for the air-annealed samples in Fig. 2. The initial (70 °C) spectrum shows strong signals from organic components in the typical C-H bond region of 2900 $cm^{-1}$, at 970 $cm^{-1}$ and 1430 $cm^{-1}$, and a strong wide band in the water and/or ethanol scattering region around 3400 $cm^{-1}$. On increasing the annealing temperature above 400 °C these bands gradually disappear and bands characteristic to the $SiO_2$ glass network increase. A sharp line at 3747 $cm^{-1}$ due to free (not H-bonded) silanol groups grows, reaches maximum at 800 °C and disappears at 1200 °C (inset of Fig. 2). The 606 $cm^{-1}$ band due to planar 3-membered rings of Si-O bonds follows roughly the same growth and decay pattern, however, does not completely vanish at 1200 °C. The final spectrum of the 1200 °C-annealed sample is very similar to that of dense type III wet silica glass (top spectrum), apart from the different concentration of H-bonded silanol groups (3690 cm$^{-1}$ band). In few cases formation of cristobalite started at 1200 °C as indicated by its sharp lines at 232 cm$^{-1}$ and 416 cm$^{-1}$ (spectra not shown).

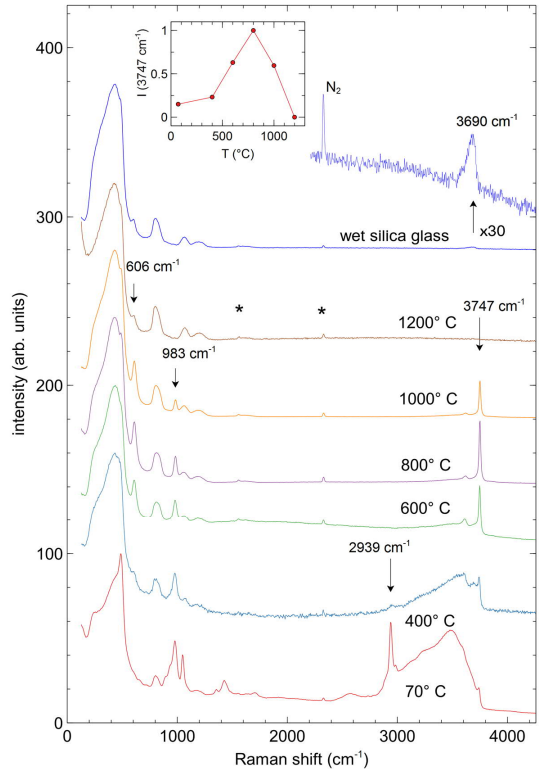

Fig. 2. Raman spectra of sample thermally annealed in air at temperatures 70 °C – 1200 °C and their comparison to the spectrum of commercial "type III" wet synthetic silica glass (top).   Inset shows the intensity dependence of the 3747 cm$^{-1}$ "free silanol" band on annealing temperature. Asterisks mark lines due to atmospheric $N_2$ (2331 cm$^{-1}$) and $O_2$ (1555 cm$^{-1}$) molecules.

### 3.2 CW-mode photoluminescence

PL emission spectra in the UV region of samples annealed in air at different temperatures are presented in Fig. 3. The excitation photon energy was chosen at 5.51 eV (225 nm) close to the maximum of the strongest PL excitation (PLE) peak (see Fig. 8 below). The most intense feature in the PL emission spectra is a composite band with an apparent peak at 3.68 eV and a shoulder due to a peak at 3.75 eV-3.80 eV. These bands emerge at 400 °C, reach maximum at ≈ 600 °C and disappear at 1200 °C.  The temperature dependence of their intensity is shown in the inset. It must be noted, however, that the sizes and shapes of differently annealed samples were different. Therefore, the intensities shown in the inset could have relative errors up to 50%, and the local minimum at 800 °C might be not meaningful (compare to the inset of the next figure, Fig.4).

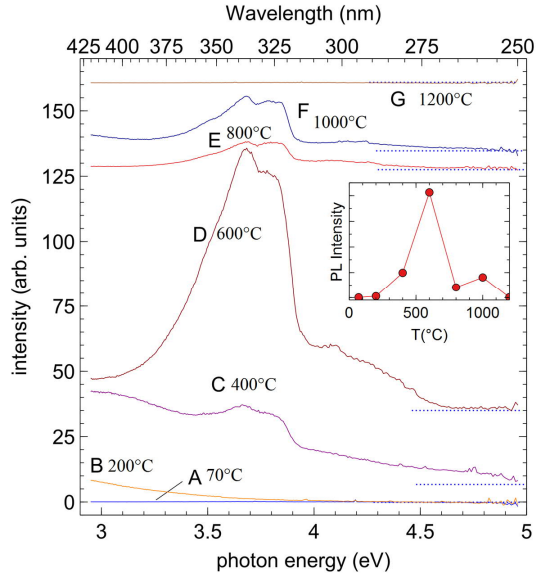

Fig. 3. UV photoluminescence spectra of samples annealed at temperatures 70 °C-1200 °C in air. Excitation photon energy 5.51 eV (225 nm). Excitation bandwidth 5 nm, PL emission spectral resolution 1 nm. Spectra are vertically shifted, baselines are indicated. Inset shows the annealing temperature dependence of the 3.68 eV PL band intensity.

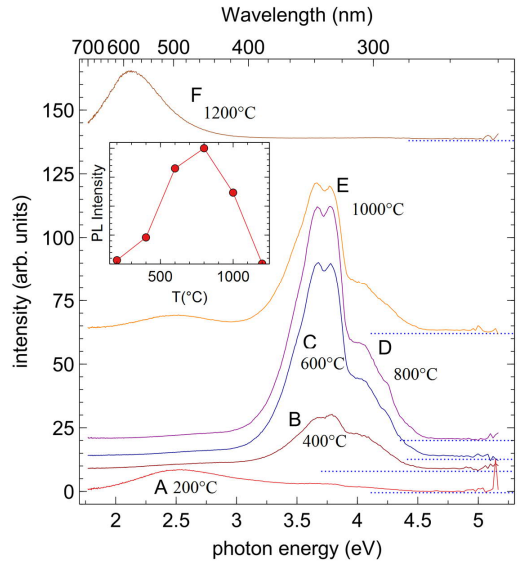

Fig. 4. PL spectra of samples annealed at temperatures 200 °C - 1200 °C in $O_2$. Excitation photon energy 5.51 eV (225 nm). Excitation bandwidth 6 nm, PL emission spectral resolution 2 nm. Spectra are vertically shifted, baselines are indicated. Inset shows the annealing temperature dependence of the 3.68 eV PL band intensity.

Figure 4. presents similar data as Fig. 3, now for the case of samples annealed in $O_2$ atmosphere. The dominant peaks are at 3.68 eV and 3.80 eV, quite similar to data of Fig. 3, both peaks are better resolved here. A further similarity, they emerge at 400 °C and vanish

at 1200 °C, the peak of the dependence of their intensity on annealing temperature (Fig. 4, inset) is shifted to the 800 °C (see above the remark on inset of Fig.3).

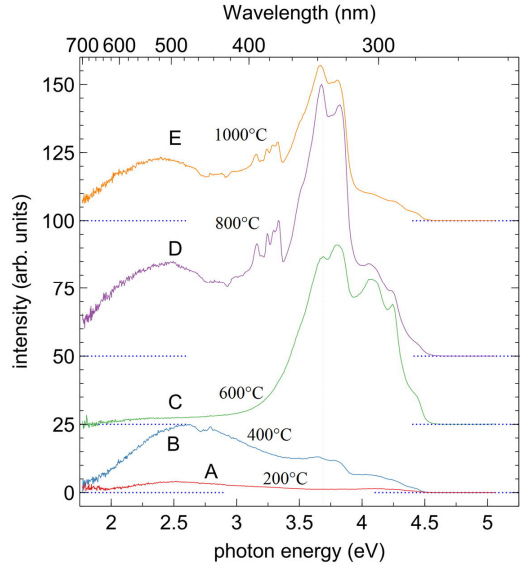

Fig.5. PL spectra of samples annealed at temperatures 200 °C- 1000 °C in $H_2$. Excitation photon energy 5.51 eV (225 nm). Excitation bandwidth 6 nm, PL emission spectral resolution 2 nm. Spectra are vertically shifted, baselines are indicated.

PL spectra of samples annealed in $H_2$ presented in Fig. 5. show some similarities to spectra in Fig. 3 and Fig. 4 and also completely new features. The same 3.68 eV and 3.80 eV PL bands, which emerge at 400 °C and vanish above 1000 °C, are present. The data on 1200 °C is absent here because the sample became strongly fumed, evidently by gas emanation at this temperature. There are new PL bands in the 4 eV – 4.5 eV range, but the most distinctive new feature is the highly structured band in the 3.1 eV – 3.3 eV range, which is created by 800 °C annealing. It becomes more prominent when time-resolved PL measurements, better spectral resolution or more appropriate excitation photon energies are used. Higher resolution spectra, recorded in smaller steps and at different excitation energies (4.66 eV and 3.73 eV) are shown below in Fig. 7A and Fig. 9A.

### 3.3 Time-resolved PL and decay kinetics

Time-resolved PL spectra measured for $H_2$-annealed sample at excitation by 4.66 eV (266 nm) photons are shown in Fig. 6. The top (black) curve in panel "A" shows spectrum, which includes both the fast and slow PL components ("quasi-CW" spectrum). It was obtained with gate width (170 µs) set longer than the period of the laser repetition rate (1/6.4 kHz=156 µs). In this way the fast components of the PL with lifetimes shorter than ≈35 ns are also included in this spectrum, because, due to instrument limitations, the acquisition starts not earlier than 35ns after the laser pulse.

The bottom 2 curves in panel "A" of Fig. 6 show PL spectra recorded with delays 35 ns and 135 ns after the laser pulse. They reveal strongly structured component in the 3 eV - 3.5 eV region and a weaker, less sharply structured component in the 3.5 eV – 4 eV region. Both these components were detected in CW-mode spectra recorded with 5.51 eV excitation (Fig. 5). The intensities of both components are markedly changed by 100 ns delay between the spectra (1.3 and 2.4 times, respectively (Fig. 6, panels "C", "B")). These graphs also illustrate that the spectral shapes do not change significantly during this time interval. Assuming mono-exponential decay law, one can calculate the decay constants as ≈ 350 ns for the narrow 3.15 eV band component and ≈110 ns for the 3.5 eV – 4 eV component.

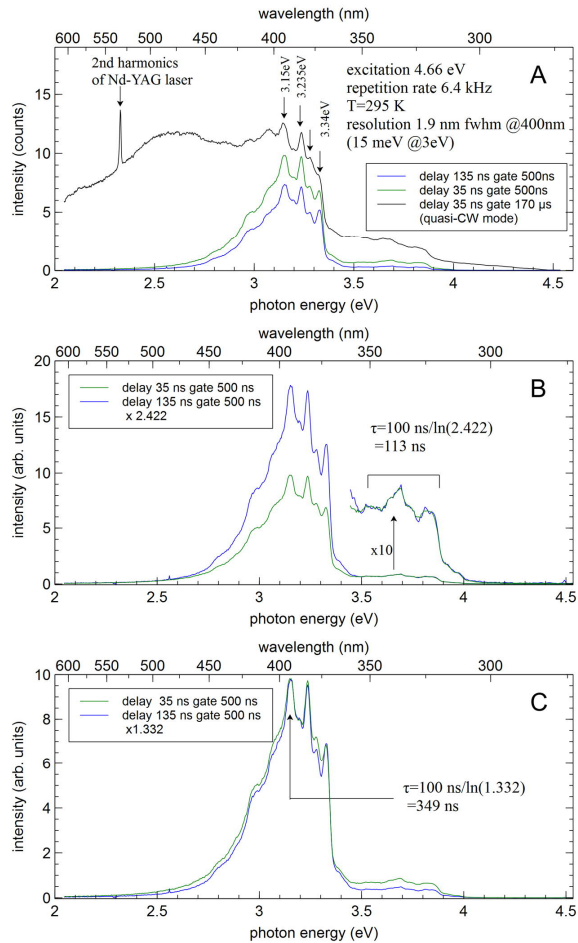

Fig. 6. Time-resolved PL spectra of sample annealed in $H_2$ at 800 °C. (A): quasi-CW mode and gated (500 ns) spectra, delayed by 35 ns and 135 ns; (B) and (C): "2-point" estimates of (assumed) exponential decay constants for the 3.5 eV – 4 eV component (B) and for the 3.15 eV line (C). Excitation energy 4.66 eV.

The kinetics of the longer-lived spectral component at 3.1 eV – 3.4 eV range is shown more accurately in Fig. 7. The top panel (A) shows the CW-mode emission spectrum, excited by 4.66 eV photons from pulsed laser operated at ~6 kHz. The decay kinetics, measured for the 3.24 eV sharp line is presented in the bottom panel (B). It was fitted with 4 exponents, the fit parameters - decay constants $\tau_i$ and intensity factors $B_i$ are indicated in the figure. About 58% of the total intensity ($\sum\tau_i B_i$) is provided by the long components with $\tau$ = 227 ns and $\tau$ = 401 ns. This finding is in accord with the estimate from time-resolved spectra measurements ($\tau$ = 349 ns), described above.

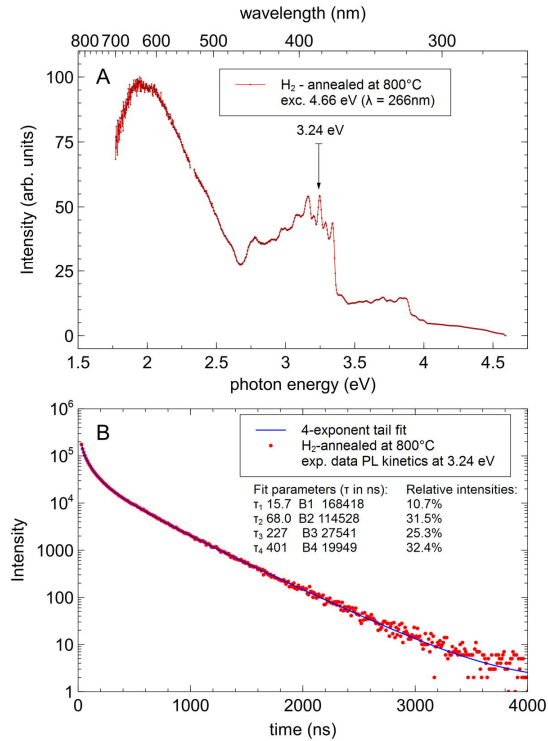

Fig. 7. Kinetics of PL in sample annealed in $H_2$ at 800 °C. Top panel (A): PL spectrum, measured in CW mode with 4.66 eV (266 nm) laser excitation and spectral resolution 1 nm. The arrow indicates the energy selected for the kinetics measurement. Bottom panel (B): PL decay kinetics measured at 3.24 eV with emission bandwidth 0.2 nm and its fit by 4 exponents.

3.4 PLE spectra

The top panel of Fig. 8 depicts PLE spectrum of the PL emission peak at 3.68 eV shown in Figs. 3 - 6. The main band at 5.64 eV (220 nm) is flanked by a structured low-energy band with a peak at 4.35 eV (285 nm) and three higher energy sub-peaks, whose mutual separations are not equal in energy scale.

Curves B and C in the middle panel show PLE spectra measured by monitoring the intensity of the narrow 3.34 eV PL band (Figs. 5 – 7) at its peak (3.34 eV B) and at its foot (3.38 eV, C). The background-corrected PLE spectrum of the sharp 3.34 eV PL band is then given by the difference spectrum D = B - C (bottom panel). The spectrum shows 3 main peaks at 3.73, 4.57 and 5.18 eV, each flanked by 2-3 higher-energy satellite peaks with uniform energy spacing of $\approx 1400$ cm$^{-1}$.

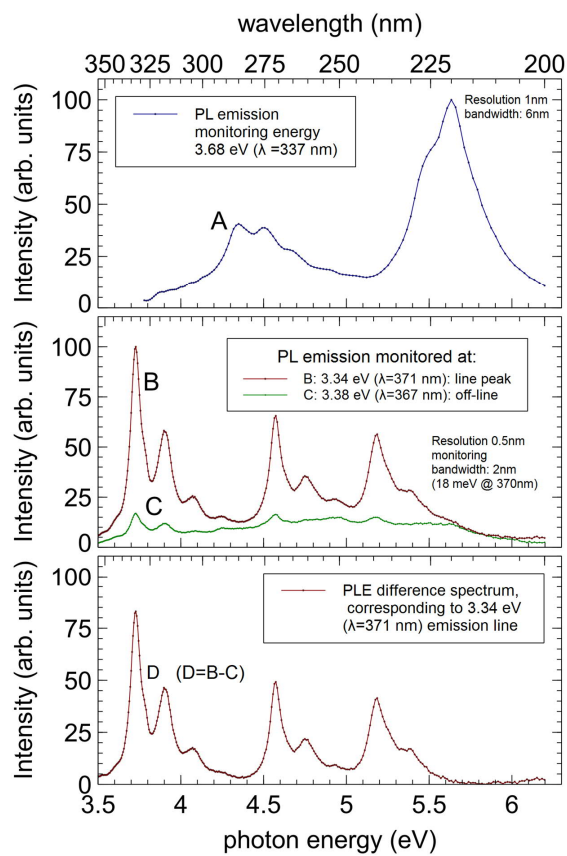

Fig. 8. Photoluminescence excitation (PLE) spectra for PL bands in sample annealed in H$_2$ at 800 °C. A: PLE of 3.68 eV PL peak (PL monitoring bandwidth 65 meV); B: PL monitored at the top of 3.34 eV peak (bandwidth 18 meV); C: PL monitored at the base of the peak, at 3.38 eV; D: difference spectrum B-C, giving the PLE spectrum of the sharp 3.34 eV peak.

# 4 Discussion

The evolution of the xerogel Raman spectra with increasing thermal annealing temperature (Fig. 2) is largely in accord with the published data [10]. The Raman spectra become similar to the spectra of dense silica glass at temperatures as low as 600 °C. The remaining differences are the increased concentration of 3-membered rings of Si-O bonds (606 cm$^{-1}$ band) and the presence of free silanol groups indicated by the 3747 cm$^{-1}$ band with its accompanying 983 cm$^{-1}$ band due to Si...(OH) stretching mode. The free silanol groups are characteristic for SiO$_2$ surfaces [11]. Their observation in samples heated up to 1000 °C (Fig. 2) indicates the presence of nano-sized internal voids.

### 4.1 The peculiarities of the observed PL

The PL emission spectra (Fig. 3 – Fig. 6) show two peculiarities, which are highly unusual for PL bands of conventional high-purity synthetic silica glasses: (i) spectral position in the near-UV range; (ii) structured spectra with sharp lines, observable at room temperature.

Despite the large number of luminescence studies of high-purity silica, apart from the well-known 4.3 eV PL band of the silicon oxygen-deficiency center ("SiODC", divalent Si) [12,13], there are much fewer reports on other UV-range PL bands, as compared to the multiple PL bands reported in the visible spectral region. Luminescence bands in the 3 eV to 4 eV region in not intentionally doped $SiO_2$ have been reported only by using other kinds of excitation, like cathodoluminescence [14], or in high-surface samples: thin films, porous materials and particles, e.g. refs. [15–20], where a possible participation of atoms other than Si, O and H in the emitting centers is possible. PL in the 3 eV – 4 eV region is reported as well in carbon-doped sol-gel glass [21] and C-implanted $SiO_2$ films [22,23].

The second peculiarity of the PL spectra in Figs. 3 - 6 and PLE spectrum in Fig. 8 is the presence of sharp vibronic lines. Intrinsic defects in silica are known commonly to undergo strong electron-phonon coupling, which causes homogeneous broadening of PL and PLE bands to FWHM larger than 0.3 eV. The static disorder by glassy state adds additional broadening of ≈0.1 eV to PL bands. This results in a typical absence of sharp zero-phonon and vibronic lines even at liquid He temperatures [12]. The only so far known exception from this rule is oxygen dangling bond "NBOHC " (non-bridging oxygen hole center), which shows zero-phonon line and vibronic structures – however, only at temperatures below 80 K, and when site-selective narrow-band excitation is applied to suppress the inhomogeneous broadening [24]. Apart from rare-earth doped glasses, all other reported cases of narrow, structured features in the PL spectra of silica pertain to interstitial molecules ($Cl_2$ [25], $S_2$ [23,26], $O_2$ [24]), or to (often unidentified) molecular adsorbates on surfaces [15,18,27–29].

Highly structured PL spectra in the 3 eV – 3.4 eV region, very similar to those observed in the present work (Fig. 5, 6, 7A), and a PLE spectrum, similar to the low energy (3.5 eV – 4 eV) part of the PLE spectrum in Fig. 8 D, have been previously reported in compressed Aerosil fumed silica nanoparticles[18,19,29]. By detailed study of dependence of PL intensities on samples type and ambient atmosphere it was shown in these studies that the intensity of this PL is proportional to the specific surface area of the sample, and that this PL is efficiently quenched by $O_2$ molecules [29].

A survey of the voluminous published work on PL of disperse or porous forms of silica reveals that the higher-energy structured PL bands found in our samples at 3.7 eV – 3.8 eV (Figs. 3 – 7) have also been previously reported in a number of papers [17,20,21,30]. The origin of these PL bands has been either tentatively related to intrinsic surface defects of silica, e.g., NBOHC's, or left open.

### 4.2 Intrinsic or extrinsic PL?

When considering the evidence on the controversy between the intrinsic or impurity-related nature of the structured PL centers in the 3 – 4 eV region, a serious "writing on the wall" is the total absence of this kind of PL in high-fluence neutron-irradiated (>$10^{19}n/cm^2$) pure silica glass, where due to the large disorder, caused by overlapping nuclear particle tracks, one could expect virtually any type of possible intrinsic (Si, O and optionally H-related) defects. An added advantage in this case is that the defects are inside the sample, free of any contamination. Due to the nanosized voids in silica glass network, both "bulk" and "surface" type defects are present. There is a long-standing observation that the surface and bulk intrinsic defects in silica tend to have similar properties [12].

The defects on surface of silica obtained under strictly controlled atmosphere have been studied in detail [31,32]. The only PL-active surface defects, reported in these studies, which have not been clearly resolved as well in particle-irradiated bulk silica glass are silanone and dioxasilirane groups - probably , because their PL bands [32] at 2.27 eV and 2.41 eV, respectively, fall into the region of other, much stronger PL bands. The situation is different in the 3 – 4 eV spectral range, which is relatively empty: neither PL of well-controlled surface defects [31] nor PL of bulk defects have been reported there. Hence, if the UV PL centers, found in the present study were intrinsic surface centers, they probably should have been detected in numerous studies of the PL of heavily particle-irradiated silica too. Their absence then most likely points to their relation to impurities, carbon-related ones being the most probable. Carbon may be introduced in sol-gel glasses by organic by-products during sol-gel synthesis. Even in the cases of $SiO_2$ synthesized without any direct contact to carbon-containing compounds, e.g., hydrophilic brands of Aerosil, large surface-area samples can adsorb ubiquitous organic molecules from air. It was recently suggested [7], that even atmospheric $CO_2$ molecules can be taken up by silica surface layer upon laser irradiation.

### 4.3 PL of naphthalene and pyrene in $SiO_2$

The most likely source of carbon in our samples is ethanol, created by hydrolysis of TEOS. Pyrolysis of ethanol is much studied in the context of engine fuel combustion, and it is known that mono- and polycyclic aromatic hydrocarbons ("PAH"s), like benzene $C_6H_6$, naphthalene $C_{10}H_8$ etc. are formed in the gas phase in the temperature region 700 °C – 1000 °C [33]. PL properties of aromatic hydrocarbons are of significant interest in many fields, dedicated textbooks [34] and spectra catalogs [35] are available. The PL spectra of the smallest aromatic hydrocarbons are (i) located in the UV spectral range (3 – 4.5 eV), and (ii) often show vibronic structures even at room temperature due to the relatively weak electron-vibrational (electron-phonon) coupling related to the delocalized character of the ground and lowest excited electronic states. These two features are present in the PL spectra of our sol-gel silica samples too. A search through an online public database [36] yields a nearly excellent agreement between the PL and PLE spectra of the sample annealed at 800 °C in $H_2$ and naphthalene molecules $C_{10}H_8$ dissolved in cyclohexane. They are plotted together in (Fig. 9); our PL spectrum there is re-measured with higher spectral resolution as compared to Fig. 5. Comparison to PL spectra of Figs. 3 and 4 show that naphthalene molecules are present also in samples annealed at 800 °C in air or $O_2$ atmosphere.

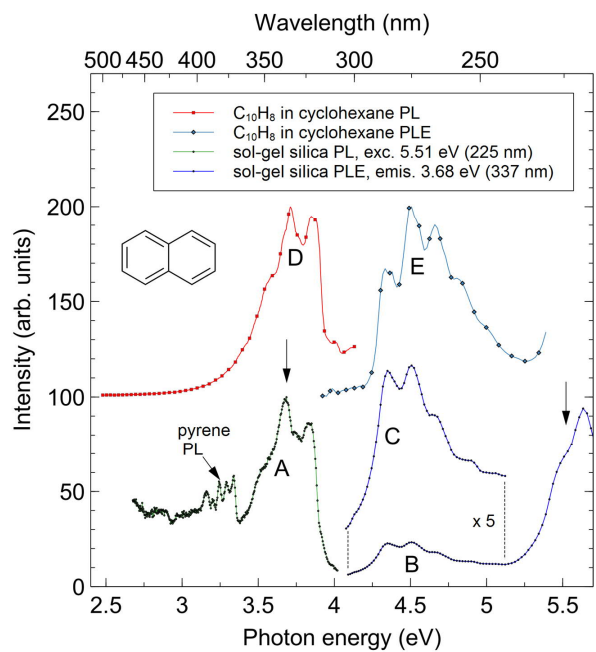

Fig. 9. Comparison of the PL (trace A) and PLE (traces B,C) spectra of sol-gel silica glass, annealed in $H_2$ at 800 °C with the published [36] PL (trace D) and optical absorption (trace E)

spectra of naphthalene molecules dissolved in cyclohexane. The vertical arrows indicate excitation (5.51 eV, bandwidth 6 nm) and emission monitoring (3.68 eV, bandwidth 6 nm) photon energies.

Spectral resolutions: 0.3 nm (PL), 1 nm (PLE)

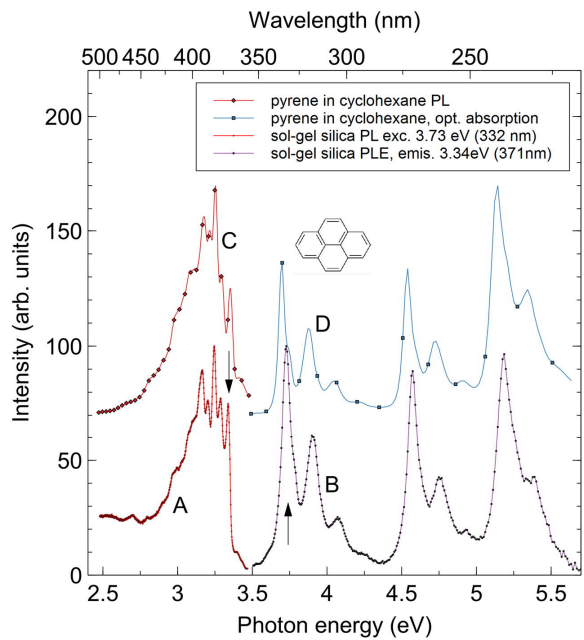

Fig. 10. Comparison of the PL (trace A) and PLE (trace B) spectra of sol-gel silica glass, annealed in $H_2$ at 800 °C with the published [36] PL and optical absorption spectra (traces C and D) of pyrene $C_{16}H_{10}$ molecules dissolved in cyclohexane. The vertical arrows indicate excitation (3.73 eV, bandwidth 6 nm) and emission monitoring (3.34 eV, bandwidth 2nm) photon energies. Spectral resolutions are 0.3 nm (PL), 0.5 nm (PLE)

When the lowest excitation peak of the 3.3 eV –3.4 eV UV PL bands (Fig. 8) is used for excitation, the resulting PL spectrum coincides very well with the PL spectrum of pyrene molecules $C_{16}H_{10}$ (Fig. 10 A, C). The relation of the PL centers in silica samples to naphthalene and pyrene molecules is further confirmed by the similarity of the shapes of PL excitation spectra with the optical absorption spectra of dissolved naphthalene and pyrene [36] (Fig. 9B, C, E, Fig. 10B, D).

Due to the highly specific spectral shapes of naphthalene and, in particular, pyrene optical bands, the identification of these PL centers in silica samples seems to be reasonably reliable. However, there are small discrepancies (≈1 – 2 nm) in spectral positions of the narrow peaks. They could be in part caused by different matrix effects (cyclohexane vs. $SiO_2$) or simply by inaccuracies in the calibration of measured and/or published spectra. Indeed, comparison of the spectra of pyrene in cyclohexane between the two databases [35,36] yields spectral shifts up to 3 nm. This introduces some uncertainty in the identification of pyrene in $SiO_2$ (Fig. 10 A, C), because there exists another 4-ring PAH, chrysene $C_{18}H_{12}$, with aromatic rings arranged in zig-zag pattern, instead of a symmetric group of 4 rings in pyrene (Fig.10). It has a structured PL spectrum too, in the same spectral region, with the main line shifted only 3 nm relative to the strongest line in pyrene spectrum.

Fortunately, pyrene and chrysene have markedly different PL decay lifetimes, reported as 390 ns and 45 ns, respectively, in deoxygenated cyclohexane [37]. The time-resolved PL (Fig. 6C) illustrates that the shape of the spectrum almost does not change on increasing the gate delay by 100 ns. The spectrum is dominated by a component with estimated lifetime of ≈350ns. It is close to the reported 390 ns lifetime of excited pyrene in cyclohexane [37]. Direct PL decay kinetics measurement (Fig. 7B) shows that components longer than 220 ns (227 ns and 401 ns) contribute to ≈ 60% of the PL intensity. The possible contribution of chrysene may be represented by the small faster-decaying component at 3.28 eV (378 nm, Fig. 6C). Its position is within error margins from the reported [35] strongest chrysene PL peak at 380 nm.

The lifetime τ=113 ns of the fast PL component at 3.5 eV – 4 eV, estimated in Fig. 6B is in accord with the lifetime of naphthalene PL in deoxygenated cyclohexane, 99 ns [37]. It must be noted, however, that the lifetimes of PAH's are sensitive to the presence of PL quenchers, usually $O_2$. The PL decay constants of naphthalene and pyrene are only 16 ns and 19 ns [37] when measured in cyclohexane containing the air-equilibrium concentration ($1.4 \times 10^{18}$ $O_2/cm^3$) of dissolved oxygen. The dependence of the lifetime of pyrene PL in Aerosil on $O_2$ and other gases has been already reported [29], with τ decreasing from 160 ns in vacuum ambient to 125 ns in 0.1 kPa $O_2$. The longer lifetimes, measured for pyrene molecules in our sol-gel glasses (230 ns and 400 ns components) show that there are not many efficient PL quenchers into their surroundings.

## 4.4 The role of polycyclic hydrocarbons in carbon-doping of silica

PL spectra in Figs. 3 – 5 are dominated by naphthalene bands. However, unresolved higher energy bands in the 4 – 4.5 eV region are also present. The work to identify them is in progress. It can be suggested that they are related to molecules with a single aromatic ring, based on the following arguments: (i) such molecules should be surely present as intermediary products in the pyrolysis reactions leading to naphthalene and pyrene formation, (ii) their expected emission energy is in the 4 - 4.5 eV spectral region (Fig. 11), in accord with the high energy shoulder of spectra in Figs. 3 – 5. Fig. 11 shows the "average PL photon energy" in dependence on the number of benzene rings in aromatic molecules. The PL energy is arbitrary represented there as an intensity-weighted average over the respective published PL spectra in databases [35,36]. This figure illustrates that single-ring molecules emit PL with the highest energies, the PL photon energy decreases with an increasing ring count, and that the PAH with a linear arrangement of rings (anthracene, tetracene, pentacene) have lower PL energies compared to other configurations with the same number of aromatic rings. Another tendency, not illustrated by Fig. 11, is that the PAH PL spectra become increasingly smoother and less structured as their hydrogens are substituted by other atoms or groups.

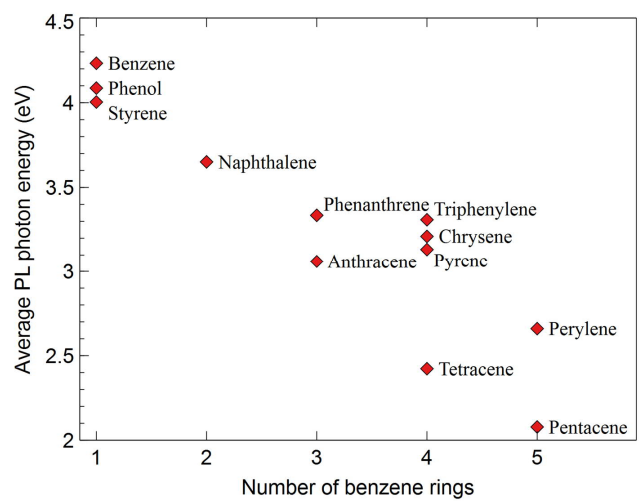

Fig. 11. Dependence on average emitted PL photon energy on the number of benzene rings in (poly)aromatic hydrocarbon (PAH) molecules. Based on PL spectra published in refs. [36] and [35] (for chrysene datapoint).

The broad spectral range of PL emissions of different aromatic hydrocarbons (Fig. 11) tempts one to extrapolate the present data and to suggest that PAHs and the structures formed after their incorporation in $SiO_2$ network could at least partially contribute to some of the numerous reported UV-VIS range PL bands in (intentionally or non-intentionally) C-doped silica samples. Data of Figs. 3, 4 indicate that PAHs in silica are relatively stable and can survive at temperatures up to 1000 °C. $^{13}$C NMR signals of PAHs are detected in silicon oxycarbide glass heated at 800 °C [5]. Ensemble of different adsorbed PAHs is demonstrated as an option to model the PL properties of carbon nanodots [38,39]. Detailed investigations of silica (Aerosil) nanoparticles covered by aromatic rings (phenyl groups) and other organic groups (propoxy, methoxy) have at first led to a suggestion that the presence of aromatic rings is not essential for PL, and that PL spectral properties are mainly determined by the crosslinking of carbon clusters to silica by Si-C and O-C bonds, where carbon is provided by a destruction of phenyl or other studied organic groups [40]. A successive study [8], however, came to a different conclusion that the "building blocks" of PL centers are aromatic rings and that the red-shift of the PL increases with the carbon ring count in the PL center. This agrees well with the hypothesis suggested by the present study.

To give more credibility to such hypothesis concerning the 3 to 4.5 eV range UV PL of silica glass, it must be explained, why the vibronic features of spectra in Figs. 3-10 of our study are present in some reported cases, e.g. [15,17,21,29,30], and absent in other cases, e.g. [8,16,20,22,23,40,41]. There can be multiple reasons: low spectral resolution (~10 nm), often used in simple PL spectrometers, an inhomogeneous broadening, overlap of multiple spectra of different PAHs, loss of vibrational features when PAH's bond to silica network - or simply that the PL origins in the second group of reports are entirely different, not related to carbon aromatic rings. This latter possibility is, however, unlikely in the case of the recent study [8] , because the PL band at 3.8 eV reported there, which is assigned to phenyl groups on silica surface, is smooth and void of any vibrational structures.

To summarize, it is well-possible that after incorporation in silica network PAHs lose some part or all of C-H bonds and form Si-C or C-O bonds resulting in small graphene-like clusters of $sp^2$-bonded C atoms, as suggested in the case of oxycarbide glasses [5]. They could give rise to a semi-continuous range of luminescence energies in the near UV –visible part of carbon-doped silica. These points are worth of further analysis.

# 5 Conclusion

The rise and decay of highly structured ultraviolet photoluminescence (PL) bands in the 3-4 eV region were observed in porous sol-gel silica glasses during their consolidation to dense glass. Based on their spectral shapes and luminescence lifetimes they can be assigned to naphthalene ($C_{10}H_8$, 2 aromatic rings) and pyrene ($C_{16}H_{10}$, 4-rings) molecules, created in the pores of glass by pyrolysis of ethanol at temperatures up to 1000 °C. Higher energy (>4eV) features in the PL spectra, indicate the presence of single-ring aromatic hydrocarbons as well. The highly specific spectral shapes of these molecules together with the high sensitivity inherent to PL methods provide a convenient tool for monitoring the sizes and incorporation of small groups of carbon atoms in silica glass. These results support the hypothesis that PL activity of carbon-doped silica may be related to carbon aromatic ring structures embedded in silica.

# Acknowledgement

The support from Latvian Science Council project lzp-2018/1-0289 and from the EU Horizon 2020 Framework Program H2020-WIDESPREAD-01-2016-2017-TeamingPhase2 under grant agreement No. 739508, project CAMART² is acknowledged.

# Appendix

## Details of sample preparation

The $SiO_2$ glass samples were prepared by the sol-gel method ("one pot synthesis") adapted from ref. [9] with modified gelation and drying process to obtain crack-free $SiO_2$ glass. The first step consisted of hydrolysis reaction at room temperature of tetraethoxysilane (TEOS, 5.58 mL, purity ≥ 99.0 %, Sigma Aldrich) in deionized water ($H_2O$, 10.0 mL), to which nitric acid ($HNO_3$, 11,16 µL, assay 70 %, Sigma Aldrich) as catalyst has been added. The resultant mixture was then stirred for 1 h at room temperature to make a homogeneous and clear solution. The molar ratio of TEOS: $H_2O$: $HNO_3$ was 1: 1.8: 0.002. The resultant homogeneous sol had pH value between 1 and 2. To increase pH value (5 - 6) in order to stop the hydrolysis reaction, ammonium acetate ($AcONH_4$, 0.1116 g, Lachner) buffer solution in $H_2O$ (45.7 mL) was prepared, and the sol was slowly added into it under constant stirring, and the resulting solution was stirred for 1 minute. Prepared sol with pH around 5-6 was then immediately divided into 6 tubes and submitted to 30 min centrifugation at 6000 rpm. After centrifugation, sols were poured into silicon molds and left for gelation at room temperature for 72 hours. Two-phase system was obtained. The liquid phase was separated and the soft gel was dried in an oven at 50 °C for 12 hours with a heating rate of 0.11 °C/min in air, followed by drying in air at 70 °C for 48 hours with a heating rate of 0.11 °C/min. The sample was then cooled down to the room temperature with a rate of 0.22 °C/min and transparent xerogel was obtained.

3 groups of dried xerogels, 6 samples each, were thermally annealed in 3 different atmospheres (air, $O_2$, $H_2$) at 6 different peak temperatures (200 °C to 1200 °C, step 200 °C), using heating rate 5 °C/minute to reach them, and then holding for 2h. After that the furnace was switched off, and samples were allowed to cool down while still keeping them in the respective atmospheres.